\documentclass[aps,prl,twocolumn,nofootinbib,showpacs]{revtex4-1}

\usepackage{bbold}
\usepackage{color}

\usepackage{amsfonts,amsmath}
\usepackage{epsfig,amsmath}
\usepackage{graphicx}
\usepackage{dcolumn}
\usepackage{bm}
\usepackage{bbold}
\newcommand{\beq}{\begin{equation}}
\newcommand{\eeq}{\end{equation}}
\newcommand{\beqa}{\begin{eqnarray}}
\newcommand{\eeqa}{\end{eqnarray}}

\newcommand{\la}{\langle} 
\newcommand{\ra}{\rangle}

\def\nat#1{{ Nat.} {\bf#1}}
\def\natphot#1{{ Nat.\ Phot.} {\bf#1}}

\def\pla#1{{ Phys.\ Lett. A\/} {\bf#1}}
\def\pra#1{{ Phys.\ Rev. A\/} {\bf#1}}

\def\prd#1{{ Phys.\ Rev. D\/} {\bf#1}}
\def\prl#1{{ Phys.\ Rev.\ Lett.} {\bf#1}}

\def\sci#1{{ Science} {\bf#1}}

\begin{document}

\title{Turning off quantum duality}
\author{X.-F. Qian$^{1,2,4}$}
\email{xiaofeng.qian@stevens.edu}
\author{K. Konthasinghe$^{1,2}$}
\author{S. K. Manikandan$^{1,3}$}
\author{D. Spiecker$^{5}$}
\author{A. N. Vamivakas$^{1,2,3}$}
\author{J. H. Eberly$^{1,2,3}$}
\affiliation{$^{1}$Center for Coherence and Quantum Optics,
University of Rochester,
Rochester, New York 14627, USA\\
$^{2}$The Institute of Optics, University of Rochester, Rochester, NY 14627, USA\\
$^{3}$Department of Physics \& Astronomy, University of Rochester,
Rochester, New York 14627, USA\\
$^{4}$Department of Physics, Stevens Institute of Technology, Hoboken, NJ 07030, USA\\
$^{5}$National Technical Institute for the Deaf, Science and Mathematics Department, Rochester Institute of Technology,
Rochester, New York 14623, USA}

\date{\today }

\begin{abstract}
We provide the first experimental confirmation of a three-way quantum coherence identity possessed by single pure-state photons. Our experimental results demonstrate that traditional wave-particle duality is specifically limited by this identity.  As a new consequence, we show that quantum duality itself can be amplified, attenuated, or turned completely off. In the Young double-slit context this quantum coherence identity is found to be directly relevant, and it supplies a rare quantitative backup for one of Bohr's philosophical pronouncements. 
\end{abstract}



\maketitle

\noindent{\bf Introduction} \quad
Louis de Broglie's hypothesis of Quantum Duality, presented in his 1924 doctoral thesis \cite{deB}, asserted that every quantum-mechanical entity will act as both a particle and a wave, two aspects equally real and contradictory.  More than 90 years later the duality hypothesis, ridiculous in ordinary discourse, is accepted by physicists as truly inescapable. 

Richard Feynman famously celebrated quantum duality  {\em ``...[as] a phenomenon which is impossible,  absolutely impossible, to explain in any classical way, and which has in it the heart of quantum mechanics. In reality it contains the only mystery."} \cite{Feynman}.  However, the mystery can be quantified, which means that duality can be tested. 

\begin{figure}[t!]
\includegraphics[width=7cm]{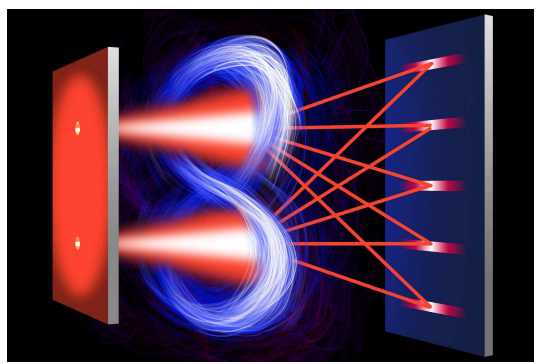}
\hspace{3cm}
\caption{Artistic conception of a generic Young-type experimental setup suggesting how entanglement engages beam distinguishability and interference visibility, all being necessarily inter-related. The source screen with the two pinholes is illuminated from the left by photons arriving singly.  } 
\label{doubleslit}
\end{figure}

Remarkably, this has never been done at the single-particle level required to test de Broglie's duality. For any experiment analogous to Young-type two-beam interference \cite{Young} (suggested in Fig. \ref{doubleslit}), the work of Wootters and Zurek in 1979 \cite{Woott-Zur} introduced the standard route for analysis and eliminated the tendency to consider duality as an exclusive either-or wave-particle restriction. They showed that it is better seen as a combination in which both features can be simultaneously active: $V$ for visibility of interference fringes measuring waveness and $D$ for separate light path distinctiveness measuring particle probability.  In several ways \cite{Greenberger-Yasin, JSV, Englert} these were then shown to be quantified in the inequality 
\beq \label{VD<1}
V^2 + D^2 \le 1.
\eeq  

Without testing duality itself, this quantification has been repeatedly exhibited and employed in various quantum scenarios  \cite{not compl, Bohr}. Examples are atom interferometry, path and phase relations of photonic orbital angular momenta, and quantum delayed-choice experiments \cite{Durr, Kurizki, Schwindt-etal, Jacques-etal, Tang-etal, Kaiser-etal, Manning-etal, Menzel, Joensuu, Bagan-etal}. The full range of both $V$ and $D$ between zero and unity has been explored, but when reporting single-particle experiments only in the specially restricted limit case $V^2 + D^2 = 1$. No single-particle experiment has tested either the inequality (\ref{VD<1}) or single-particle duality more generally. 

One must remember that there is another mystery that is as deeply embedded within quantum theory as duality. This is entanglement, known for its mysterious hands-off control of Schr\"odinger's hypothetical Cat. A recent classical optical analysis by three of us \cite{QVE} (QVE for short) suggested that entanglement is a coherence missing from  (\ref{VD<1}). This does not contradict the Wootters-Zurek finding. Concurrence $C$ \cite{Wootters} is recognized as the appropriate entanglement measure in a qubit-like (two-beam) interference context, and the QVE analysis suggests that concurrence  enters single-particle duality directly because it completes the inequality (\ref{VD<1}) to an equality, a quadratic identity for two-beam interference:
\beq \label{VDC=1}
V^2 + D^2 + C^2 = 1. 
\eeq

By convention (see Feynman \cite{Feynman}), duality refers to properties of a single physical entity. This is also our focus here, although we point out that a similar inequality has been derived for the case of two-qubits \cite{Jak-Berg}.  Our report is about the first experimental observations that test the quantification of duality. Our test is compatible with the presence of perfectly coherent single-particle entanglement via concurrence. Moreover, we explain in our concluding remarks that identity (\ref{VDC=1}) allows us a rare experimental confirmation. We provide quantitative support for one of the many semi-philosophical pronouncements that Bohr was known for, and which are almost never open to direct test. 

The prior QVE analysis is not fully compelling because it was based on classical optical physics, and the reported experiments relied on entirely non-quantum instrumentation and detection. However, the classical optical derivation did take advantage of a mathematical inseparability.  As Schr\"odinger noted when he introduced the term ``entanglement" to quantum theory \cite{Schr}, such inseparability is the signal of entanglement. This is true whether the states being entangled are vectors in quantum Hilbert space or in classical linear function space. Thus the question is, does the three-part classical identity (\ref{VDC=1}) derived in QVE also apply to individual quantum particles? In the following paragraphs we present the results of fully quantum single-particle (photon) self-interference experiments that give a positive answer to the question, and thus represent the first completely tested quantification of quantum duality. 

\begin{figure}[h!]
\includegraphics[width=6cm]{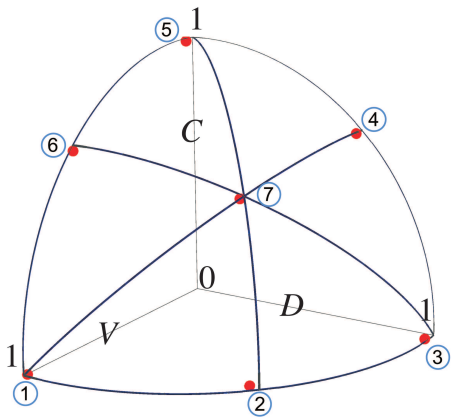}
\caption{An octant is shown of the VDC sphere defined by the QVE duality identity $V^2 + D^2 + C^2 = 1$. Our experiments subject a generic single-photon pure state to a variety of different two-beam interference experiments. These yield a wide range of $V$ and $D$ values. The observed values of $V^2 + D^2$ vary from 0 to 1, but in all cases remain on the surface of the $VDC$ sphere, consistent with (\ref{VDC=1}).}
\label{VDCsphere}
\end{figure}

\noindent{\bf Experimental background} \quad
Here we present the results of our experiments, made with individual pure-state photons that create Young-type two-beam, or two-way Mach-Zehnder, self-interference. To begin the quantum experimental discussion, we repeat that de Broglie's duality is understood as an inescapable part of the nature of every single electron, photon, neutron, atom, etc. Thus, in the following we stick to the original discussions of single-particle duality and do not consider coherences of multiple particles (see, e.g., \cite{Jak-Berg, Rio}).

The details of a two-beam interference experiment (recall Wootters and Zurek \cite{Woott-Zur})  can be arranged for a given input photon pure state to yield any value of $V$ and any value of $D$ consistent with the often-proved inequality $V^2 + D^2 \le 1$. Obviously, when the single photon is completely a particle ($D=1$) then there should be no wave character ($V=0$), and vice versa.  However, relation (\ref{VD<1}) doesn't really predict a balanced exchange between $V$ and $D$ simply because the inequality permits a decrease of $D$ and $V$ together, or an increase by both. It even allows the extreme decrease $V$=$D$=0 to exist (neither wave nor particle - see Sperling, et al. \cite{Sperling}) while there is still a photon on hand. Such a photon can't be nothing. One sees that something must be missing from $V^2 + D^2  \le1 $.\

Our task is to examine experimentally the proposition that $V$ and $D$ are restricted in a way that the inequality  doesn't address. That is, the QVE condition $V^2 + D^2 + C^2 = 1$ is proposed as the correct general  single-particle duality restriction.  To proceed, the generic single-photon pure state to be tested is given by 
\beq
|\Psi\ra = c_a|\mathbb{1}_a\ra \otimes |\phi_a\ra + c_b|\mathbb{1}_b\ra \otimes |\phi_b\ra, 
\label{partial-coherent}
\eeq
where $c_a$, $c_b$ are normalized coefficients with $|c_a|^2+|c_b|^2=1$; and $|\mathbb{1}_a\ra$, $ |\mathbb{1}_b\ra$ are single photon mode states indicating respectively one photon in propagating modes $a$, $b$ and no photon elsewhere. Here $|\phi_a\ra$ and $|\phi_b\ra$ are two corresponding normalized states of all the remaining intrinsic (usually continuous) degrees of freedom (e.g., polarization, temporal mode, etc.) of the single photon with correlation defined as $|\gamma| = |\la \phi_a|\phi_b\ra|\le1$. From (\ref{partial-coherent}) and the expression for $\gamma$, one can quickly retrieve  inequality (\ref{VD<1}) in the form
\beq \label{VD}
V^2 + D^2 = 1 - 4|c_a^*c_b|^2(1 - |\gamma|^2) \le 1,
\eeq
which our experiments show to be incomplete.\\

\begin{figure*}[t!]
\includegraphics[width=18cm]{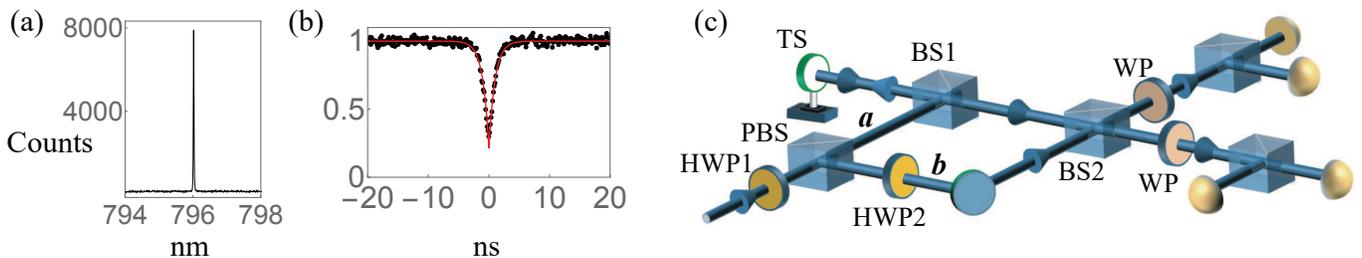}
\caption{ Experimental aspects in testing the three-part single-particle duality identity $V^2 + D^2 + C^2 = 1$ with hBN-generated individual entangled pure-state photons defined by (\ref{AEexp}). Details of parts (a)-(c) are described in the text.}
\label{bigfig}
\end{figure*}

\noindent{\bf Single photon generation} \quad
We employ a defect-hosted monolayer hexagonal boron nitride (hBN) quantum emitter to serve as a single photon source. Studies of a two-dimensional hBN quantum emitter as a quantum light source were started in 2015 by Tran, et al. \cite{Tran-etal}.  In this Letter we make the first systematic use of such hBN-generated photons in the realization and measurement of self interference and single-photon entanglement.  Our two-dimensional hBN sample resides on a silicon substrate ($\approx$ 5mm$\times$5mm in size) which is mounted (on an X-Y-Z Piezo) in an Attocube cryostat maintained at 10K. A Ti:sapphire laser operated at 725nm is directed to the sample through a confocal microscope to produce off-resonant excitations of an hBN quantum dot. The emission spectrum, illustrated in Fig.~\ref{bigfig} (a), shows a bright zero-phonon line centered at 796.03nm with a linewidth $\le0.2$ nm (limited by the instrument resolution 0.2 nm). The narrow linewidth promises good coherence with single photon interference visibility up to $98.3\%$ realized in the measurement stage. The second-order correlation function $g^{(2)}(\tau)$ of the detected signal is measured through a Brown-Twiss setup. Due to the detection resolution limit, the minimum value measured is $g^{(2)}(0)=0.29$ indicating single photon nature, as shown in Fig. \ref{bigfig}(b). The detailed properties of the hBN photon source are described in Ref. \cite{Konthasinghe-etal}.\\

\noindent{\bf Single-photon entangled state preparation} \quad
We prepared seven different single photon states based on (\ref{partial-coherent}), for testing the completed three-way QVE identity $V^2 + D^2 + C^2 = 1$ . We employed a Mach-Zehnder interferometer (MZI), an analog of Young's double slits, to create the path states $|\mathbb{1}_a\ra$, $|\mathbb{1}_b\ra$. For practical consideration and without loss of generality, the states of all remaining degrees of freedom of the photon $|\phi_a\ra$, $|\phi_b\ra$, are represented by the polarization states $|s_a\ra$, $|s_b\ra$. 

The hBN quantum dot emitted single photon is spectrally filtered with a 10-nm bandpass filter and polarization-oriented with a horizontal polarizer $|h\ra$.  As shown in Fig. \ref{bigfig}(c), the polarized polarized single photon passes through a half-wave plate (HWP1) and changes into an arbitrarily polarized state $|s\ra$. A polarizing beam splitter (PBS) separates the horizontal $|h\ra$ and vertical $|v\ra$ polarizations into paths $a$ and $b$ respectively. The transmitted component, $|\mathbb{1}_a\ra \otimes |h\ra$, in path $a$ passes through a 50/50 beamsplitter (BS1) that directs to a mirror mounted on a translation stage, before heading to the second 50/50 beamsplitter (BS2). The reflected component of PBS in path $b$ passes through a half-wave plate (HWP2) to become $|\mathbb{1}_b\ra \otimes |s_b\ra$ where $|s_b\ra = e^{i\xi}\cos\theta |h\ra + \sin\theta|v\ra$. Then the single photon state entering the combining beamsplitter (BS2) can be described as
\beq \label{AEexp}
|\Psi\ra = c_a|\mathbb{1}_a\ra \otimes |s_a\ra + c_b|\mathbb{1}_b\ra \otimes |s_b\ra,
\eeq
with $|s_a\ra=|h\ra$ representing horizontal polarization. Here the amplitude ratio $R=|c_b/c_a|$ of the two components is controlled by HWP1. The overlap of the two polarization states can be characterized as $\gamma = \la s_a|s_b\ra = \cos\theta e^{i\xi}$, where $\theta$ is controlled HWP2 in path $b$ and $\xi$ is manipulated by the translation stage in path $a$. 

According to the quadratic QVE identity (\ref{VDC=1}), the values of $(V, D, C)$ occupy the positive octant of the VDC sphere. We chose seven representative sets of $V,D,C$ distributed over this octant surface. These target sets are determined respectively by the nodes of six grid lines on the $VDC$ sphere, i.e., $V=0$, $D=0$, $C=0$, $V/D=1$, $V/C=1$, and $C/D=1$, as shown in Fig.\ref{VDCsphere}. The detailed values of all sets ($V,D,C$) are given in detail in the Table below.

We achieved these target $(V, D, C)$ values by generating corresponding single photon states as defined in Eq.~(\ref{AEexp}), by way of the state control parameters $R$ and $\theta$. This is based on the fact that visibility, distinguishability as well as entanglement concurrence (between the path states $|\mathbb{1}_a\ra$, $|\mathbb{1}_b\ra$ and polarization states $|s_a\ra$, $|s_b\ra$) can be expressed in terms of the parameters $R$ and $\theta$, i.e., 
\beq \label{DV-parameter}
 V=\frac{2R|\cos\theta |}{1+R^2}, \quad D=\left| \frac{1- R^2}{1+R^2}\right|,  \quad C=\frac{2R|\sin\theta |}{1+R^2}.
\eeq  \\

\noindent {\bf Confirming Results} \quad For a given single-photon state, fringe visibility $V$ is achieved by registering the photon counts with an APD at the output of BS2 while continuously moving the translation stage. The particle distinguishability $D$ is obtained straightforwardly with photon counts by blocking first one and then the other of the two paths. Measurement of concurrence $C$ is realized by a tomographic analysis \cite{James-etal}. We measured $V$, $D$, $C$ values for all seven single-photon states and the results are presented in Table \ref{Table}. They are also illustrated on the VDC sphere in Fig.~\ref{VDCsphere}, where the circled numbers correspond to the state numbers in the table. 

\begin{table}[t!] 
\begin{center}
\begin{tabular}{|r|r|r|r|r|r|rl}
\hline
& \ $V\ \  $ &$D\ \  $&$C\ \  $& $V^2+D^2 $&$ SUM \simeq 1\ \ $ \\
\hline 
\hspace{0.5mm}
1&\ 0.992 &\  0.009 &\  0.003 &\ 0.985\ \ \  &\ \ 0.985 $\pm$ 0.014\ \  \\
\hline 
\hspace{0.5mm}
2\ & 0.719  & 0.680 & 0.012 & 0.980\ \ \    & 0.980 $\pm$ 0.054\ \ \\
\hline 
\hspace{0.5mm}
3\ & 0.068 & 0.994 & 0.008 & 0.992\ \ \    & 0.992 $\pm$ 0.060\ \   \\
\hline 
\hspace{0.5mm}
4\ & 0.048  & 0.708 & 0.703 & 0.503\ \ \    & 0.998 $\pm$ 0.084\ \ \\
\hline 
\hspace{0.5mm}
5\ & 0.058 & 0.011 & 0.991 & 0.004\ \ \    & 0.986 $\pm$ 0.040\ \  \\
\hline 
\hspace{0.5mm}
6\ & 0.720 & 0.011 & 0.691 &0.518\ \ \    & 0.996 $\pm$ 0.070\ \ \\
\hline 
\hspace{0.5mm}
7\ & 0.587 & 0.568 & 0.570 &0.667\ \ \   & 0.992 $\pm$ 0.070\ \  \\
\hline 
\end{tabular}
\end{center}
\caption{The measured values of visibility, distinguishability, and concurrence, where ${\rm SUM}$ refers to the identity combination $ V^2 + D^2 + C^2 $ and state 5 is an example supporting our remark that even lossless pure states can have the traditional quantity of duality completely turned off - i.e., $ V^2 + D^2 \simeq 0$.} \label{Table}
\end{table}

\begin{figure}[h!]
\includegraphics[width=6.2cm]{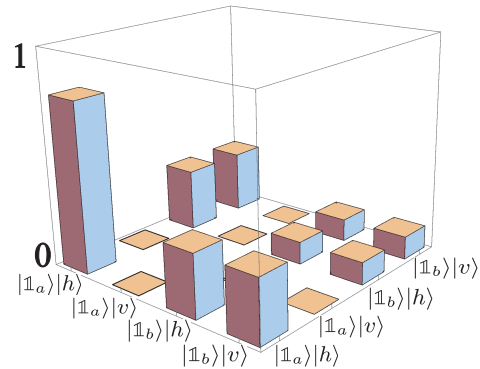}
\caption{Measured density matrix of state $|\Psi\ra$ in Eq.~(\ref{Equal}).} 
\label{densmatrix}
\end{figure}

The table's SUM column reports $V^2 + D^2 + C^2 \simeq 1$ for each one-photon state, \#1 to \#7. Those values represent confirmation of identity (\ref{VDC=1}). As an example, one special ``equal coherence" case of the three-way identity is achieved with $V=D=C$, indicating matching contributions of wave, particle, and entanglement. The corresponding one-photon self-entangled pure state is given as 
\beq \label{Equal}
|\Psi\ra = \sqrt{\frac{3+\sqrt{3}}{6}}|\mathbb{1}_a\ra |h\ra + \sqrt{\frac{3-\sqrt{3}}{12}}|\mathbb{1}_b\ra (|h\ra+|v\ra).
\eeq
The density matrix of the tomographically measured experimental state (\ref{Equal}) is displayed in Fig. \ref{densmatrix} \\

\noindent {\bf Concluding Remarks:}  \quad We observed with single-photon detection the QVE three-way identity $V^2 + D^2 + C^2 = 1$, for the quantum coherence possessed by single photons. At least since the 1980s, treatments of duality have been quantified in terms of  $V$ and $D$ wave-particle analysis. This is shown to be incomplete. In the Young double-slit context the new identity (\ref{VDC=1}) among $V$, $D$, and $C$ is seen to supplant it, establishing the important role for self-entanglement (non-separability of degrees of freedom) in the self-interference of individual quantum particles. 

Finally, there is relevance of these results even to the normally untestable Principle of Complementarity. Bohr's 1949 summary of his mandate (see  \cite{moreBohr}) reveals this point. As he said ``...  {\em evidence obtained under different conditions cannot be comprehended within a single picture, but must be regarded as complementary in the sense that only the totality of the phenomena exhausts the possible information about the objects}". Entanglement is a coherence that we believe was never considered by Bohr, and by taking account of single-particle self-entanglement our experiment supplies what seems to be the first quantitative backup for his personal insights. Our results show experimentally that the repeatedly derived and employed wave-particle inequality $V^2 + D^2 \le 1$ does not exhaust the possible information about a single quantum (recall the Wooters-Zurek analysis, and a speculation by Knight in 1998 \cite{Knight98}). The completed QVE identity $V^2 + D^2 + C^2 = 1$, which adds single-particle entanglement, does so and says that by including self-entanglement all possible information of a single quantum object is exhausted in a two-slit experiment. In effect, we are reporting the experimental observation of the exhaustion of the coherences of single photons.  \\

\noindent\textbf{Acknowledgement:} \quad
We are pleased to acknowledge conversations with colleagues G.S. Agarwal, I. Bialynicki-Birula, Paul Brumer, Steven Cundiff, ¬†F. De Zela, ¬†B.-G. Englert, A. Friberg, Peter Knight, Peter Milonni, Michael Raymer, L. Sanchez-Soto, W.P. Schleich, ¬†J.P. Torres, W.H. Zurek and financial support from DARPA D19AP00042, and NSF grants PHY-1203931, PHY-1505189, and INSPIRE PHY-1539859.


\begin{thebibliography}{99}
\bibitem{not compl} We keep in mind the lack of a firm relation between duality and complementarity. The former is de Broglie's simple flat statement about each single quantum object, and complementarity (notably not mentioned by Feynman) is Bohr's phrasing of the consequences of duality. Bohr's multiple rephrasings are well known - see \cite{Bohr}.
\end{thebibliography}
\end{document}